\documentclass[prl,twocolumn,twoside,preprintnumbers,superscriptaddress,floatfix]{revtex4}

\usepackage{amsmath}
\usepackage{amsfonts}
\usepackage{amssymb}
\usepackage{color}
\usepackage{enumitem}
\usepackage{float}
\usepackage{graphicx}
\usepackage{hyperref}
\usepackage{multirow}
\usepackage{setspace}
\usepackage{ulem}
\usepackage{xcolor}
\usepackage{xspace}
\usepackage{ulem}

\def\be{\begin{equation}}
\def\ee{\end{equation}}
\def\beq{\begin{equation}\begin{aligned}}
\def\eeq{\end{aligned}\end{equation}}

\newcommand{\Eq}[1]{Eq.~\eqref{#1}}

\newcommand{\Ref}[1]{Ref.~\cite{#1}}

\begin{document}

\title{Sub-GeV Dark Matter Shining at Future MeV Gamma-Ray Telescopes}

\author{Francesco~D'Eramo}
\email{francesco.deramo@pd.infn.it}
\affiliation{Dipartimento di Fisica ed Astronomia, Universit\`a di Padova, Via Marzolo 8, 35131 Padova, Italy}
\affiliation{INFN, Sezione di Padova, Via Marzolo 8, 35131 Padova, Italy}

\author{Stefano~Profumo}
\email{profumo@ucsc.edu}
\affiliation{Department of Physics, 1156 High St., University of California Santa Cruz, Santa Cruz, CA 95064, USA}
\affiliation{Santa Cruz Institute for Particle Physics, 1156 High St., Santa Cruz, CA 95064, USA}

\date{\today}

\begin{abstract}
We propose a novel framework where light (sub-GeV) dark matter (DM) is detectable with future MeV gamma-ray telescopes without conflicting with Cosmic Microwave Background (CMB) data. The stable DM particle $\chi$ has a very low thermal relic abundance due to its large pair-annihilation cross section. The DM number density is stored in a slightly heavier, meta-stable partner $\psi$ with suppressed pair-annihilation rates, that does not perturb the CMB, and whose late-time decays $\psi \rightarrow \chi$ fill the universe with $\chi$ DM particles. We provide explicit, model-independent realizations for this framework, and discuss constraints on late-time decays, and thus on parameters of this setup, from CMB, Big Bang Nucleosynthesis, and Large Scale Structure. 
\end{abstract}

\maketitle

\noindent 
{\bf Introduction.} Models of dark matter (DM) with sub-GeV thermal relics are, generically, strongly constrained by Cosmic Microwave Background (CMB) data. A DM particle with a thermal relic density matching the measured cosmological DM abundance corresponds to a total annihilation cross section times the relative velocity at freeze-out of~\cite{Lee:1977ua,Scherrer:1985zt,Srednicki:1988ce,Gondolo:1990dk,Steigman:2012nb}
\be
\langle \sigma v_{\rm rel} \rangle_{\rm F.O.}  \simeq 
3 \times 10^{-26} \, \text{cm}^3 \, \text{sec}^{-1} \ .
\label{eq:sigmavrelTH}
\ee
Here, the product $\sigma v_{\rm rel}$ is averaged over a thermal ensemble. CMB forms when the universe is $\tau_{\rm CMB} \simeq 380,000$ years old (the time of ``recombination''), typically long after DM freeze-out. Nevertheless, the energy injected by out-of-equilibrium DM annihilation can measurably alter the CMB anisotropy spectrum~\cite{Chen:2003gz,Padmanabhan:2005es,Galli:2009zc,Slatyer:2009yq,Finkbeiner:2011dx}. Quantitatively, CMB data constrain the DM pair-annihilation rate at recombination to be ~\cite{Ade:2015xua}
\be
\langle \sigma v_{\rm rel} \rangle_{\rm CMB} \lesssim  \frac{4.1 \times 10^{-28} \, {\rm cm}^3 {\rm s}^{-1}}{f_{\rm eff}} \, \left(\frac{m_{\rm DM}}{{\rm GeV}} \right) \ .
\label{eq:sigmavrelCMB}
\ee
The ``efficiency factor'' $f_{\rm eff}$ is approximately constant for a given annihilation channel, and the specific values can be found in Refs.~\cite{Slatyer:2015jla,Galli:2011rz,Giesen:2012rp}. For typical values $f_{\rm eff} \simeq 0.5$, thermal relics with $m_{\rm DM} \lesssim 35 \, {\rm GeV}$ are excluded. 

Well-known caveats to the conclusion above exist: The annihilation rate can decrease as the universe expands, such as for p-wave or forbidden annihilations~\cite{Griest:1990kh,DAgnolo:2015ujb}; one could modify the Hubble expansion rate by injecting entropy after freeze-out~\cite{McDonald:1989jd,Kamionkowski:1990ni,Giudice:2000ex,Co:2015pka,Hamdan:2017psw}, allowing for smaller annihilation rates; the relic density could  be set by a primordial asymmetry~\cite{Nussinov:1985xr,Kaplan:1991ah,Kaplan:2009ag}, by co-scattering~\cite{DAgnolo:2017dbv}, by co-annihilations~\cite{DAgnolo:2018wcn}, or by $3 \rightarrow 2$ processes~\cite{Carlson:1992fn,Hochberg:2014dra}. While these possibilities all evade CMB bounds, they also weaken indirect detection signals today, in most cases making the indirect detection of sub-GeV DM very challenging. The DM annihilation rate always satisfies the bound in \Eq{eq:sigmavrelCMB}, which is well below the current~\cite{Essig:2013goa} and future~\cite{Boggs:2006mh,2012ExA,Hunter:2013wla,Wu:2014tya,Tatischeff:2016ykb} sensitivity of gamma-ray telescopes.  One possible exception is a large annihilation rate in the co-decaying scenario of Ref.~\cite{Dror:2016rxc}; here, however, CMB limits are evaded only if there is a very high degree of mass degeneracy in the particle spectrum in the dark sector, producing a velocity suppression to the annihilation rate at recombination.

In this Letter, we propose a novel framework for sub-GeV DM with large indirect-detection signals in the late universe: The DM candidate is a stable particle $\chi$ that annihilates to Standard Model (SM) final states through fast, s-wave processes. We add to the dark sector (at least) a new, heavier metastable degree of freedom $\psi$, charged under the same symmetry ensuring the stability of the $\chi$ DM particle, and with a lifetime longer than $\tau_{\rm CMB}$. The DM particle is coupled to the thermal bath at high temperatures, and its large annihilation rate significantly washes out $\chi$'s at freeze-out. The dark matter number is stored into the heavier partner $\psi$,  stable until after recombination, and whose interactions do not deposit any energy in the electron-photon plasma.  After the last scattering surface epoch, $\psi \rightarrow \chi$ decays populate the universe with DM particles: the universe today is thus filled with stable $\chi$'s that have a large annihilation cross section, producing significant and possibly detectable signals for future MeV gamma-ray telescopes. 

The DM relic density as ``measured'' by CMB~\cite{Ade:2015xua} is reproduced for $\rho_{\rm DM} /s = (m_\chi n_\chi + m_\psi n_\psi) / s \simeq 0.44 \, {\rm eV}$, where $s$ is the entropy density. Regardless of how the relic density is shared between $\chi$ and $\psi$, the CMB bound has to be rescaled, here, to account for the fact that only $\chi$'s annihilations affect recombination. After presenting two general scenarios realizing our framework, with a detailed discussion of relic density and CMB bounds, we discuss constraints on late time decays and annihilations in the ``cosmic dark ages'', i.e. after recombination, and how they affect the parameter space of our framework.

\begin{figure}
\centering
\includegraphics[width=0.99\linewidth]{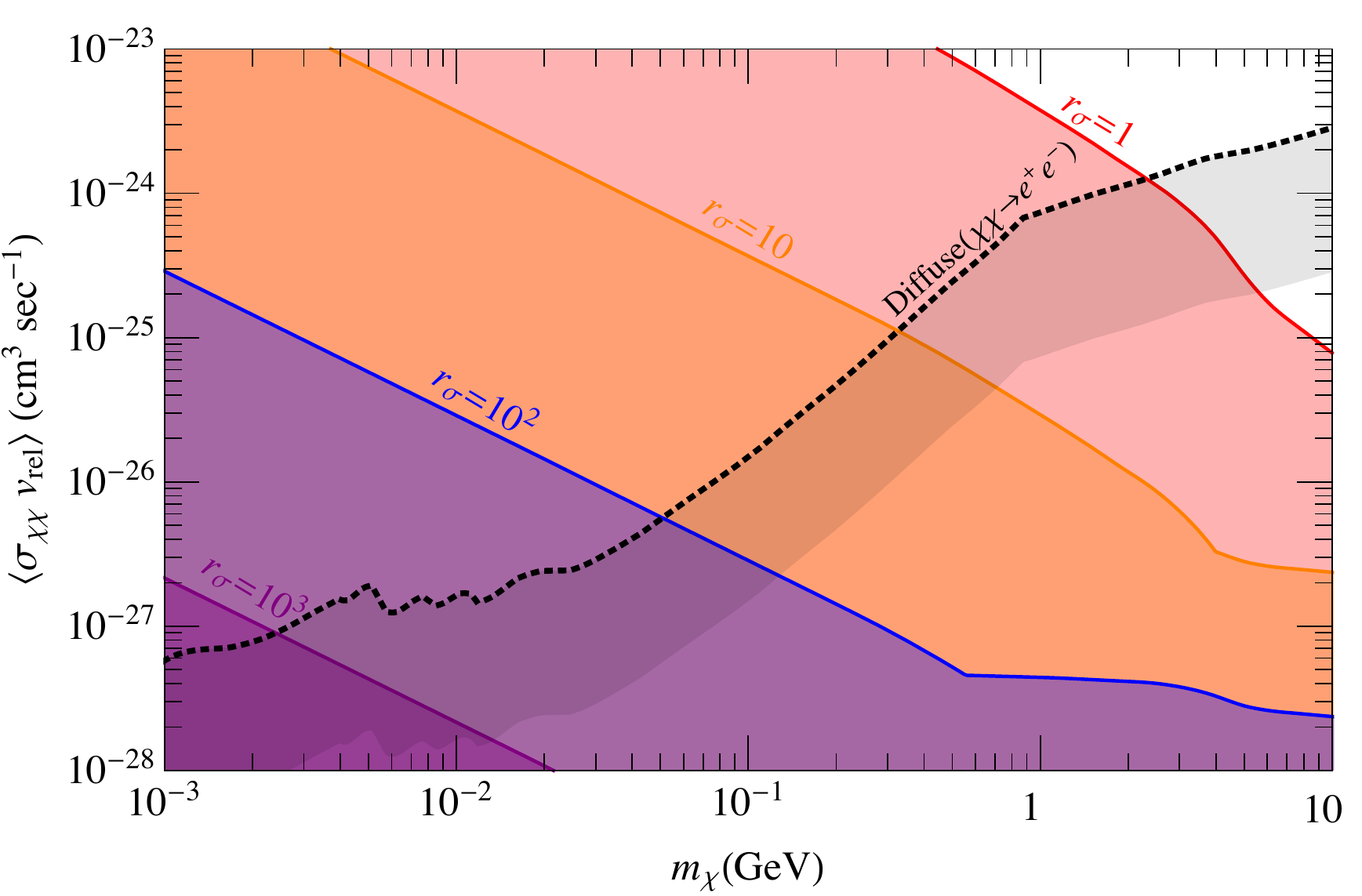}
\caption{Allowed parameter space for $\psi$ non-thermal production. For different values of $r_\sigma$ (see main text), we shade regions excluded by overproduction and/or CMB bounds. The region below each curve is excluded. We also show current diffuse limits for $\chi \chi \rightarrow e^+ e^-$.}
\label{fig:NonThermal}
\end{figure}

\noindent 
{\bf Two Scenarios.} We first consider a weakly coupled partner $\psi$ produced non-thermally. Freeze-out of $\chi$'s is not affected by the presence of $\psi$, and number densities can be tracked by solving the Boltzmann system
\begin{align}
\label{eq:BEFI1} 
\frac{d n_\chi}{dt} + 3 H n_\chi = & \, - \langle \sigma_{\chi \chi} v_{\rm rel} \rangle_{\text{\tiny (F.O.)}} 
\left[n_\chi^2 - n_\chi^{{\rm eq} \, 2}\right]  \ , \\
\label{eq:BEFI2} 
\frac{d n_\psi}{dt} + 3 H n_\psi =  & \, \mathcal{C}_\psi  \ ,
\end{align}
where $H$ is the Hubble parameter describing the expansion rate of the universe. We do not include operators accounting for the late time decays $\psi \rightarrow \chi$, since they happen long after freeze-out. In particular, $\psi$'s decay after recombination, and in order to properly rescale CMB bounds we need to know the {\it individual} number density of $\chi$ {\it and} $\psi$. The two differential equations are decoupled and can be solved independently. The most minimal option would be to populate the early universe with $\psi$'s via $\chi \rightarrow \psi$ inverse decays. This would not require new interactions, since the coupling inducing $\psi \rightarrow \chi$ must exist to mediate late-time decays. However, such a coupling is bound to give a lifetime longer than the time of CMB formation,  $\tau_\psi > \tau_{\rm CMB}$. The comoving density of $\psi$ particles produced via this mechanism is approximately $\rho_\psi / s \simeq M_{\rm Pl} / (m_\psi \tau_\psi)$. The relic density constraint combined with the bound on the lifetime of $\psi$ implies $m_\psi \lesssim 0.3 \, {\rm eV}$, way beyond the range of our investigation. A mechanism that would work is freeze-in production~\cite{Hall:2009bx} through decays of a heavier particle $\Phi$, corresponding to $\mathcal{C}_\psi = n_\Phi^{\rm eq} \,  \Gamma_{\Phi \rightarrow \psi} \, K_1[m_\Phi/T] / K_2[m_\Phi/T]$. Here, we do not specify the source of non-thermal production $\mathcal{C}_\psi$, our findings are general.
 
We show the available parameter space for this first scenario in the $(m_\chi, \langle \sigma_{\chi\chi} v_{\rm rel} \rangle)$ plane in Fig.~\ref{fig:NonThermal}. The variable on the vertical axis is the s-wave annihilation cross section to {\it visible channels} and {\it evaluated today}, which is the observable probed by indirect searches. The effective cross section at freeze-out appearing in \Eq{eq:BEFI1} does not necessarily have to be the same; there could be annihilation to invisible or forbidden channels, co-annihilations, etc. We quantify this mismatch by introducing the parameter
\be
r_\sigma \equiv \frac{\langle \sigma_{\chi \chi} v_{\rm rel} \rangle_{\text{\tiny (F.O.)}} }{\langle \sigma_{\chi \chi} v_{\rm rel} \rangle} \ .
\label{eq:rsigmadef}
\ee
We consider the representative benchmark $f_{\rm eff} = 0.5$ for the CMB bound, and we always take Weyl fermions ($g_\chi = g_\psi = 2$; note that this choice does not impact our findings). For four different values of $r_\sigma$, we shade regions that are not phenomenologically viable. At small annihilation cross sections, one typically overproduces $\chi$'s. For larger annihilation cross sections, instead, one typically underproduces $\chi$'s, and the measured DM abundance can be accounted for by production of $\psi$'s. Whatever mechanism we employ to produce $\psi$, as long as it gives the correct DM abundance, the rescaled CMB bounds open up a significant parameter space region. Values for $r_\sigma > 1$ further widen the available parameter space. The result does not depend on the specific choice of $m_\psi$. The boundary of the excluded region for each $r_\sigma$ is made of two pieces, which can understood by using a semi-analytical solution~\cite{Kolb:1990vq} for $\chi$ freeze-out abundance. The roughly horizontal piece of the boundary (only visible for the blue and orange regions) corresponds to overproduction of $\chi$, and it is approximately given by
\be
\begin{split}
\langle \sigma_{\chi \chi} v_{\rm rel} \rangle  \simeq & \, 5.2 \, \times \, 10^{-24} \, {\rm cm}^3 {\rm sec}^{-1} \, 
\frac{1}{r_\sigma} \\ & \left( \frac{x_{f(\chi)}}{20} \right) \left( \frac{10}{g_*(x_{f(\chi)})} \right)^{1/2}   \ .
\end{split}
\ee
Here, $x_{f(\chi)} = T_{f(\chi)} / m_\chi \simeq 20$ is the freeze-out temperature and $g_*(x_{f(\chi)})$ is the number of relativistic degrees of freedom at that time. The portion of the boundary with negative slope corresponds to the rescaled CMB bound, which approximately results in
\be
\begin{split}
\langle \sigma_{\chi \chi} v_{\rm rel} \rangle \gtrsim & \; 3.3 \times 10^{-24}  \, {\rm cm}^3 {\rm sec}^{-1} \,  \frac{1}{r_\sigma^2}   \left( \frac{1 \, {\rm GeV}}{m_\chi} \right) \times \\ & \left(\frac{f_{\rm eff}}{0.5}\right) \left( \frac{x_{f\chi}}{20} \right)^2 \left(\frac{10}{g_*(x_{f\chi})}\right)   \ .
\end{split}
\label{eq:CMBboundGeneral}
\ee
For reference, we also show, shaded in grey, the region of parameter space amenable to indirect searches, which is capped by a dashed black line indicating constraints from diffuse photon emission for DM annihilation to electron/positron pairs~\cite{Essig:2013goa}. As noticeable from Fig.~\ref{fig:NonThermal}, our framework in this first scenario provides sub-GeV DM candidates well within the reach of future gamma ray missions.

A second, different possibility is if the heavier partner originates from thermal freeze-out, and thus originally was in thermal equilibrium, as long as its annihilation rate is sufficiently suppressed at recombination. The Boltzmann system for this case reads
\begin{align}
\label{eq:BEFO1} 
\frac{d n_\chi}{dt} + 3 H n_\chi = & \, - \langle \sigma_{\chi \chi} v_{\rm rel} \rangle_{\text{\tiny (F.O.)}} \left[n_\chi^2 - n_\chi^{{\rm eq} \, 2}\right] 
\\ & \nonumber + \langle \sigma_{\psi \rightarrow \chi} v_{\rm rel} \rangle_{\text{\tiny (F.O.)}} 
\left[n_\psi^2 - r_{\rm eq}^2 n_\chi^2\right] \ , \\
\label{eq:BEFO2} 
\frac{d n_\psi}{dt} + 3 H n_\psi =  & \, - \langle \sigma_{\psi \psi} v_{\rm rel} \rangle_{\text{\tiny (F.O.)}}  \left[n_\psi^2 - n_\psi^{{\rm eq} \, 2}\right] + \\ & \nonumber
- \langle \sigma_{\psi \rightarrow \chi} v_{\rm rel} \rangle_{\text{\tiny (F.O.)}}  \left[n_\psi^2 - r_{\rm eq}^2 n_\chi^2\right]  \ . 
\end{align}
For simplicity, we assume that annihilations are s- and p-wave processes for $\chi$ and $\psi$, respectively. Thus we parameterize 
\begin{align}
\sigma_{\chi \chi} v_{\rm rel} = & \, \sigma_s = \langle \sigma_{\chi \chi} v_{\rm rel} \rangle_{\text{\tiny (F.O.)}} \ , \\
\sigma_{\psi \psi} v_{\rm rel} = & \, \sigma_p v_{\rm rel}^2 \ .
\end{align}
The ratio between the effective annihilation cross section at freeze-out for $\chi$ and the one to visible final states today is here, again, given by the factor $r_\sigma$ defined in \Eq{eq:rsigmadef}. We also introduce the ratio between the p- and s-wave freeze-out cross section parameters
\be
r_\beta \equiv \frac{\sigma_p}{\sigma_s} \ . 
\ee
A similar two-state system with a long-lived partner was considered in a different context by Ref.~\cite{Fairbairn:2008fb}, for weak scale DM and only accounting for annihilations. Here, we also include species conversion reactions $\chi \chi \leftrightarrow \psi \psi$ since they can affect relative as well as overall number densities (see e.g. \Ref{DEramo:2010keq}). In both equations, we use detailed balance to always write the collision operator for the reaction $\psi \psi \rightarrow \chi \chi$, which is the one allowed at zero kinetic energy, and we define the ratio between equilibrium number densities
\be
r_{\rm eq} \equiv \frac{n_\psi^{{\rm eq}}}{n_\chi^{{\rm eq}}} \ .
\ee
As shown in the next section due to CMB limits, the ratio $m_\psi / m_\chi$ has to be very close to an integer number, up to few percent. For the calculation of abundances from freeze out, we can ignore this small deviation and only consider integer mass ratios. In order to account for the phase-space suppression, we parameterize species conversion cross sections as follows
\begin{align}
\sigma_{\psi \rightarrow \chi} v_{\rm rel} = & \, \left\{ 
\begin{array}{ccccccl} 
\kappa \, v_{\rm rel}  & & & & & & m_\psi / m_\chi \simeq 1  \\
\kappa  & & & & & & m_\psi / m_\chi \simeq 2, 3, \ldots
\end{array} \right. \ .
\label{eq:sigmaconv}
\end{align}
The effect of this type of reactions is quantified by the dimensionless parameter
\be
r_\kappa \equiv \frac{\kappa}{\sigma_s} \ .
\ee

\begin{figure}
\centering
\includegraphics[width=0.99\linewidth]{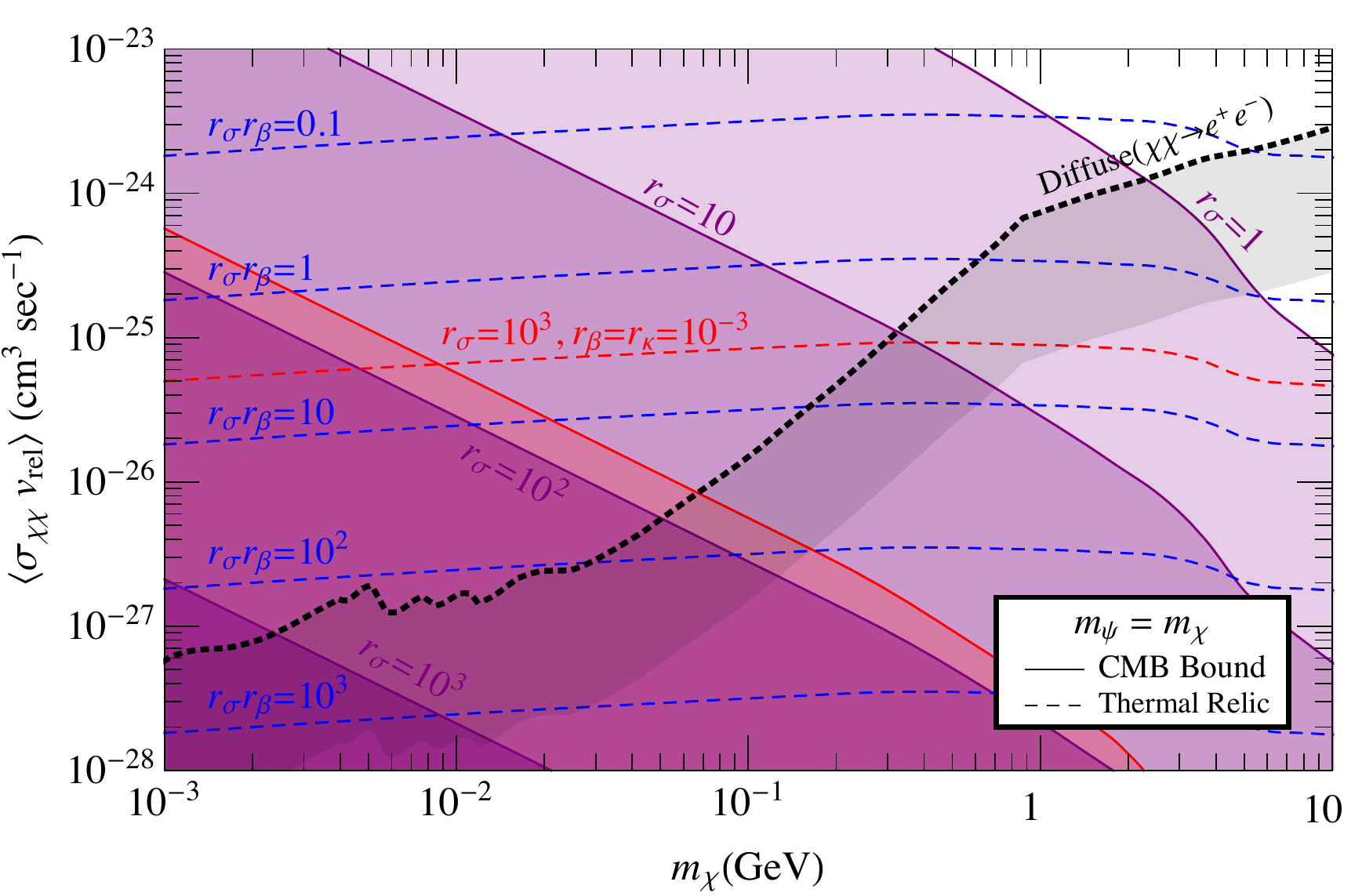}
\caption{Parameter space for frozen-out $\psi$. Dashed blue lines reproduced the observed DM density for each value of $r_\sigma r_\beta$ (see main text) and without species conversions. Always in the absence of species conversion, purple region are excluded by CMB bounds for each $r_\sigma$. We present results for one case where species conversions are present, with relic density and CMB constraints (dashed and solid red lines, respectively). As in Fig.~\ref{fig:NonThermal}, we show current diffuse limits for $\chi \chi \rightarrow e^+ e^-$.}
\label{fig:Thermal}
\end{figure}

Results for this second scenario are shown in Fig.~\ref{fig:Thermal}, where we present the available parameter space always in the $(m_\chi, \langle \sigma_{\chi\chi} v_{\rm rel} \rangle)$ plane. We fix $m_\chi \simeq m_\psi$, and start our exploration by neglecting species conversions (i.e. $r_\kappa = 0$), thus $\chi$ and $\psi$ freeze-out {\it independently}. It is possible in this case to derive semi-analytical results for the relic density and CMB bounds. We are interested in regions of parameter space giving mostly $\psi$ production at freeze out, in order to relax CMB bounds. With this assumption, we employ again a known semi-analytical solution~\cite{Kolb:1990vq} for the freeze-out density, and find the following condition to obtain the correct relic density:
\be
\begin{split}
\langle \sigma_{\chi \chi} v_{\rm rel} \rangle  \simeq & \, 2.1 \, \times \, 10^{-24} \, {\rm cm}^3 {\rm sec}^{-1} \, 
\frac{1}{r_\sigma r_\beta} \\ & \left( \frac{x_{f(\psi)}}{20} \right)^2 \left( \frac{10}{g_*(x_{f(\psi)})} \right)^{1/2}   \ .
\end{split}
\ee
This closely corresponds to the blue dashed lines in Fig.~\ref{fig:Thermal}, which were obtained by a full numerical solution of the Boltzmann system in Eqs.~\eqref{eq:BEFO1} and \eqref{eq:BEFO2}. The rescaled CMB bounds are approximated by the inequality
\be
\begin{split}
\langle \sigma_{\chi \chi} v_{\rm rel} \rangle \gtrsim  & \; 3.3 \, \times \, 10^{-24} \, {\rm cm}^3 {\rm sec}^{-1} \,  \frac{1}{r_\sigma^2} \left(\frac{1 \, {\rm GeV}}{m_\chi} \right) 
\, \times \\ & \left( \frac{f_{\rm eff}}{0.5} \right)\,   \left( \frac{x_{f(\chi)}}{20} \right)^2 \left( \frac{10}{g_*(x_{f(\chi)})} \right) \ ,
\end{split}
\ee
and this constraint corresponds to the excluded purple regions in Fig.~\ref{fig:Thermal}, obtained again by a full numerical solution. The effect of species conversions cannot be captured by simple analytical expressions since $\chi$ and $\psi$ do not freeze out independently anymore. We choose a specific set of values for the parameters and we show numerical results for this case in Fig.~\ref{fig:Thermal}  with red lines. As long as the parameter $r_\kappa$ is small, the effect is not dramatic and the overall picture is unchanged. However, for $r_\kappa \simeq 1$ the early universe can no longer be filled mostly with $\psi$'s particles: the species conversion reaction $\psi \psi \rightarrow \chi \chi$ becomes an effective way to wash the universe out of $\psi$'s, and we end up with comparable abundances of $\chi$ and $\psi$. As a result, our mechanism works as long as $r_\kappa \lesssim 10^{-1}$. We note however that, alternatively, one can invoke a form for the species conversion cross section different from the one in \Eq{eq:sigmaconv}, as for example one including further velocity suppression, effectively weakening the value of $r_\kappa$. The figure also shows the region covered by future indirect searches. Our framework predicts candidates within the reach of these experiments, even if the heavier partner $\psi$ frozen-out originally was in thermal equilibrium and its relic abundance was set by the standard freeze-out process.

{\bf Cosmological constraints from the ``dark ages''.} Thus far we have been agnostic about the mass splitting between the $\psi$ and $\chi$ particles, $\Delta m=m_\psi-m_\chi$, and the precise value for the $\psi$ lifetime. Here we use cosmological constraints to describe the viable space for these parameters in the present framework. We find that constraints from the Big Bang nucleosynthesis of light elements (BBN) at most force $\Delta m/m_\chi\lesssim 10^{-5}$ \cite{Cyburt:2002uv, Cyburt:2009pg, Poulin:2015opa}, a value much less stringent than those from distortion of the CMB spectrum after recombination from late-time decays to visible-sector particles.

Decays after CMB recombination lead to significant constraints from effects on the CMB anisotropy spectrum. Ref.~\cite{Slatyer:2016qyl}  calculates the maximum allowed fraction $f_{\rm dec}$ of decaying dark matter with a given lifetime $\tau_\psi$, for a broad range of spectra and injected energies. To be as conservative as possible, we use the strongest limits on such fraction (i.e. the smallest value of $f_{\rm dec}$ from Fig.~11 of Ref.~\cite{Slatyer:2016qyl}) to constrain the quantity  $\Delta m/m_\chi\simeq\Delta m/m_\psi<f_{\rm dec}$ at different lifetimes.

After production, $\chi$ annihilations also contribute to perturbing observables related to the CMB, and are constrained by limits similar to those described above and given in Eq.~(\ref{eq:sigmavrelCMB}), with an important difference. Neglecting the effects of clumping due to structure formation~\cite{Finkbeiner:2011dx}, integrating in redshift Eq.~(14) of Ref.~\cite{Slatyer:2012yq} in the limit of large $\tau_\psi$, and assuming that the CMB limits on $\langle\sigma_{\chi\chi} v_{\rm rel}\rangle$ scale linearly with the injected energy per unit volume, the limits are weakened by the factor $\left(z_{\rm dec}(\tau_\psi)/z_{\rm rec}\right)^3$~\footnote{A more thorough calculation including e.g. the effects of redshift-dependent energy absorption efficiency and ionization is beyond the scope of the present study.}. Here, $z_{\rm rec} \simeq 1100$ is the recombination redshift and $z_{\rm dec}(\tau_\psi)$ is the redshift corresponding to the decaying particle lifetime $\tau_\psi$. The value for the allowed lifetime range is thus found as as follows
\be
z_{\rm dec}(\tau_\psi)\simeq z_{\rm rec}\left(\frac{8\times 10^{-28} \frac{{\rm cm}^3 \, {\rm sec}^{-1}}{ {\rm GeV}}}{\frac{\langle \sigma_{\chi \chi} v_{\rm rel}  \rangle}{m_\chi}}\right)^{1/3} \ .
\ee
We consider here three values for the quantity 
\be
\frac{\langle \sigma_{\chi \chi} v_{\rm rel} 
\rangle}{m_\chi} = \left\{ 10^{-26},\ 10^{-24},\ 10^{-22} \right\}  \frac{{\rm cm}^3 \, {\rm sec}^{-1}}{ {\rm GeV}} \ ,
\ee
which correspond to the following redshifts and ages/lifetimes:
\be
z_{\rm dec} = \left\{ 474,\  102,\ 22 \right\} \ ,
\ee
\be
\frac{\tau_\psi}{\rm sec}= \left\{ 4.7\times 10^{13},\ 5.1\times 10^{14},\ 5.0\times 10^{15} \right\} \ .
\label{eq:taupsi}
\ee
No constraints exist for parameters corresponding to $z_{\rm dec}>z_{\rm rec}$, i.e. $\langle \sigma_{\chi \chi} v_{\rm rel} \rangle/m_\chi\lesssim 8\times 10^{-28}\frac{{\rm cm}^3}{\rm s\ GeV}$

Finally, while low-redshift experiments indicate a smaller value of $\Omega_m$ than high-redshift experiments such as Planck (see e.g. Fig.~11 of Ref.~\cite{Abbott:2017wau}), the discrepancy is at present within 1 standard deviation, and therefore does not constrain $\Delta m$ in our scenario. It is worth noticing, though, that the general trend of shrinking $\Omega_m$ and $\sigma_8$ from high to low redshift is an expected outcome of models such as those discussed here (for related discussions in the context of decaying dark matter see e.g. Ref.~\cite{Berezhiani:2015yta,Enqvist:2015ara}).

\begin{figure}
\centering
\includegraphics[width=0.95\linewidth]{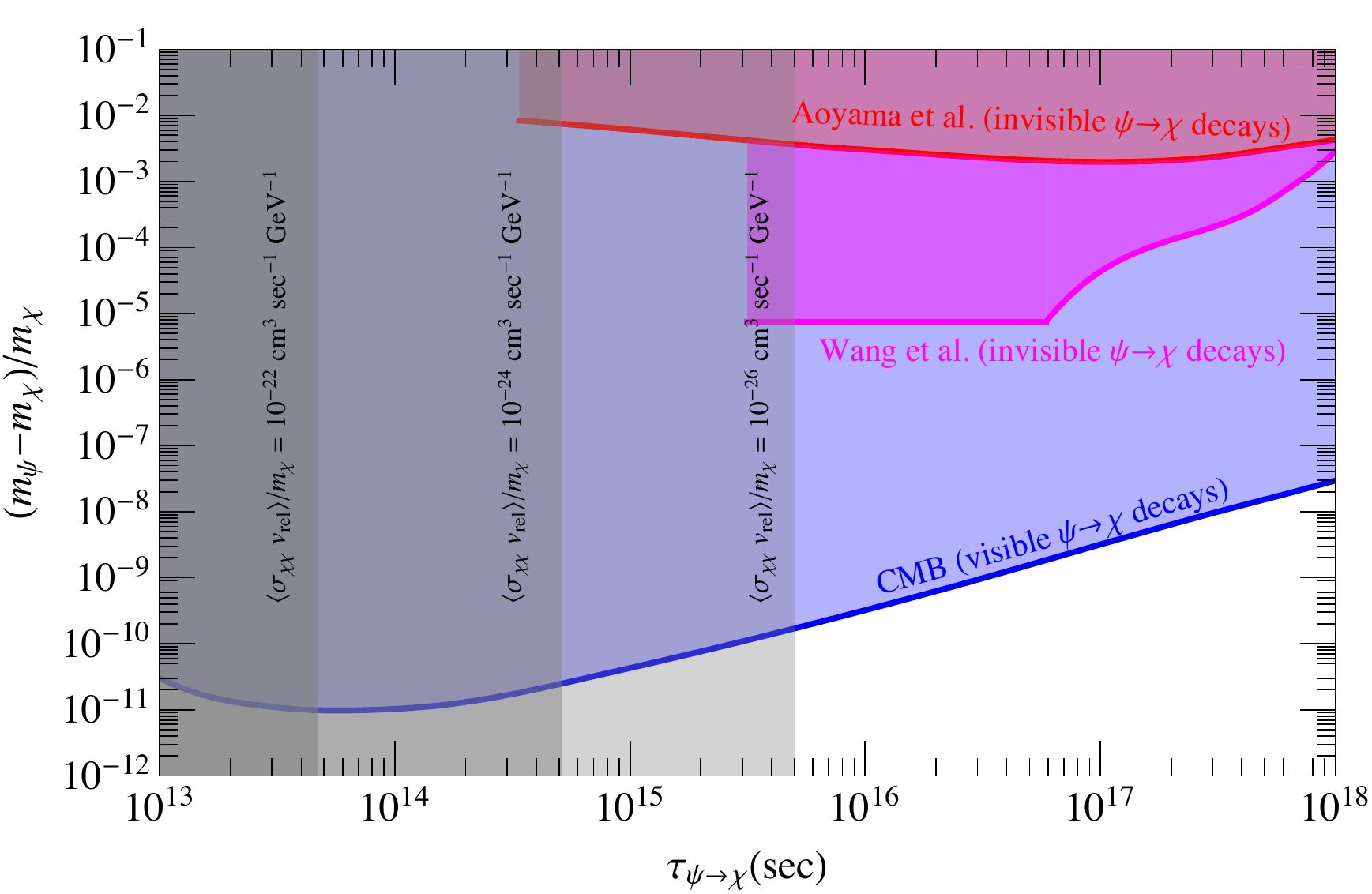}
\caption{Model independent constraints on the lifetime versus relative mass splitting plane. Parameter space above the blue line and shaded in blue is ruled out by constraints from CMB on $\psi$ decays; finally, parameter space to the left of the three vertical lines is ruled out by the late-time annihilation of $\chi$ particles for $\langle \sigma_{\chi \chi} v_{\rm rel} \rangle/m_\chi=10^{-26},\ 10^{-24},\ 10^{-22}\  \frac{{\rm cm}^3}{\rm s\ GeV}$, left to right. We also show constraints for the case of invisible decay final products from Refs.~\cite{Wang:2013rha} and \cite{Aoyama:2014tga} (the lines and shaded areas correspond to the lifetime ranges considered in those analyses).}
\label{fig:modelindep}
\end{figure}

In Fig~\ref{fig:modelindep}, we summarize these model-independent constraints. For visible $\psi\to\chi$ decays, the region at large $\Delta m/m_\chi$ above the blue line is ruled out by constraints from post-recombination decays while the regions to the left of the vertical lines indicate constraints from annihilations. The level of degeneracy in the $\psi$, $\chi$ mass spectrum required to prevent spoiling CMB data is thus very high, typically of order of one part in $10^9$. We notice that CMB constraints could be however significantly relaxed for very small mass splitting, $\Delta m\ll 1$ eV, since the energy of the final state particles would be below the hydrogen binding energy and they would not be able to modify CMB formation. 

If the daughter particle(s) in the $\psi\to\chi$ decay is (are) invisible, such as for example SM neutrinos of new light degrees of freedom, the constraints just discussed do not apply, and the only relevant constraints arise from the effect of the late decays on the formation of structures; such effects have been studied in detail with N-body simulations in Ref.~\cite{Wang:2013rha,Cheng:2015dga,Buch:2016jjp} and for a broader range of lifetimes, including $\tau\ll1$ Gyr, in Ref.~\cite{Aoyama:2014tga}; we show such constraints in Fig.~\ref{fig:modelindep}; for short lifetimes (still much larger than the epoch of recombination), the relative mass splitting can be as large as a percent.

\noindent 
{\bf Discussion and Conclusions.} In this Letter, we considered the possibility of light, sub-GeV dark matter $\chi$ annihilating at large rates today while evading CMB and other cosmological constraints. The key idea is that the dark matter is produced from late-time, post-recombination decays of a slightly heavier particle $\psi$ whose annihilation does not significantly perturb the CMB. We solved in detail the coupled Boltzmann equations describing the evolution of the number densities of the two species $\chi$ and $\psi$, as a function of their respective annihilation rate and of the strength of the operator(s) connecting their number densities. We calculated conservative constraints from annihilations of relic $\chi$'s and from the decays themselves, and argued that it is generically possible to have light, sub-GeV dark matter with large late-time annihilation rates compatibly with CMB constraints.

Our findings motivate the physics case for utilizing future satellite missions~\cite{Boggs:2006mh,2012ExA,Hunter:2013wla,Wu:2014tya,Tatischeff:2016ykb} aimed at exploring the gamma-ray sky in the energy range between $0.1 \, {\rm MeV}$ up to $100 \, {\rm MeV}$, for dark matter searches. They also motivate future studies of more complete and independently motivated particle physics constructions based on late-time, post-recombination dark matter production from decays.

\medskip
{\it Acknowledgments.}--- We thank Eric Kuflik, Yonit Hochberg, Alessio Notari, Maxim Pospelov, Yael Shadmi and Tracy Slatyer for useful discussions. FD was supported by Istituto Nazionale di Fisica Nucleare (INFN) through the ``Theoretical Astroparticle Physics'' (TAsP) project. SP was supported in part by the U.S. Department of Energy grant number DE-SC0010107.

\bibliographystyle{apsrev}
\bibliography{SubGeVindirect}

\end{document}